\documentstyle[twoside,fleqn,espcrc2]{article}
\newcommand{\Lienard}{Li\'enard}
\newcommand{\Barcelo}{Barcel\'o}

\newcommand{\etc}{\emph{etc}}

\newcommand{\eg}{\emph{e.g.}}
\newcommand{\beq}{\begin{equation}}
\newcommand{\eeq}{\end{equation}}
\newcommand{\beqn}{\begin{eqnarray}}
\newcommand{\eeqn}{\end{eqnarray}}

\newcommand{\AmS}{{\protect\the\textfont2
  A\kern-.1667em\lower.5ex\hbox{M}\kern-.125emS}}
\title{Superluminal censorship} 
\author{Matt Visser\address{Physics Department, Washington University,
            Saint Louis, Missouri 63130-4899, USA},
       B.A. Bassett\address{Relativity and Cosmology Group,
           Portsmouth University,~PO1~2EG, U.K.}
	   \thanks{Department of Theoretical Physics,
           University of Oxford, 1 Keble Road, OX1 3NP, U.K.},
        S. Liberati\address{International School for Advanced Studies 
           (SISSA), Via Beirut 2--4, 34014 Trieste, Italy}
           \thanks{Istituto Nazionale di Fisica Nucleare (INFN),
            sezione di Trieste, Italy}
}

\begin{document}

\begin{abstract}
We argue that ``effective'' superluminal travel, potentially caused by
the tipping over of light cones in Einstein gravity, is always
associated with violations of the null energy condition (NEC).  This
is most easily seen by working perturbatively around Minkowski
spacetime, where we use linearized Einstein gravity to show that the
NEC forces the light cones to contract (narrow). Given the NEC, the
Shapiro time delay in any weak gravitational field is always a delay
relative to the Minkowski background, and never an
advance. Furthermore, any object travelling within the lightcones of
the weak gravitational field is similarly delayed with respect to the
minimum traversal time possible in the background Minkowski geometry.
\end{abstract}

\maketitle

\section{INTRODUCTION}

The relationship between the causal aspects of spacetime and the
stress-energy of the matter that generates the geometry is a deep and
subtle one. In this note we report on the perturbative investigation
of the connection between the null energy condition (NEC) and the
light-cone structure. We shall demonstrate that in linearized gravity
the NEC always forces the light cones to contract (narrow): Thus the
validity of the NEC for ordinary matter implies that in weak
gravitational fields the Shapiro time delay is always a delay rather
than an advance.

This simple observation has implications for the physics of
(effective) faster-than-light (FTL) travel via ``warp drive''. It is
well established, via a number of rigorous theorems, that any
possibility of effective FTL travel via traversable wormholes
necessarily involves NEC
violations~\cite{Morris-Thorne,MTY,Visser,HV}.  On the other hand, for
effective FTL travel via warp drive (for example, via the Alcubierre
warp bubble~\cite{Alcubierre}, or the Krasnikov FTL
hyper-tube~\cite{Krasnikov}) NEC violations are observed in specific
examples but it is difficult to prove a really general theorem
guaranteeing that FTL travel implies NEC violations. A number of
partial results are known, and it is clear that at least part of the
problem arises in even defining what we mean by FTL. Recent progress
in this regard has been made by Olum~\cite{Olum}.

In this note we shall largely restrict attention to weak gravitational
fields and work perturbatively around flat Minkowski spacetime.  One
advantage of doing so is that the background Minkowski spacetime
provides an unambiguous definition of FTL travel. A second advantage
is that the linearized Einstein equations are simply (if formally)
solved via the gravitational {\Lienard}--Wiechert potentials. The
resulting expression for the metric perturbation provides information
about the manner in which light cones are perturbed.

\section{LINEARIZED GRAVITY}

For a weak gravitational field, linearized around flat Minkowski
spacetime, we can in the usual fashion write the metric
as~\cite{Visser,MTW,Wald}
\begin{equation}
g_{\mu\nu} = \eta_{\mu\nu} + h_{\mu\nu},
\end{equation}
with $h_{\mu\nu} \ll 1$. Then adopting the Hilbert--Lorentz gauge
(\emph{aka} Einstein gauge, harmonic gauge, de Donder gauge, Fock
gauge)
\begin{equation}
\partial_\nu \left[ h^{\mu\nu} - {1\over2} \eta^{\mu\nu} h \right] = 0,
\end{equation}
the linearized Einstein equations are~\cite{Visser,MTW,Wald}
\begin{equation}
\Delta h_{\mu\nu} = 
-16\pi G \left[ T_{\mu\nu} - {1\over2} \eta_{\mu\nu} T \right].
\end{equation}
In terms of the trace-reversed stress tensor
\begin{equation}
\overline{T}^{\mu\nu}(\vec y,\tilde t) =  
T_{\mu\nu}(\vec y,\tilde t) - 
{1\over2} \eta_{\mu\nu} T(\vec y,\tilde t),
\end{equation}
this has the formal solution~\cite{Visser,MTW,Wald}
\begin{equation}
h_{\mu\nu}(\vec x,t) =
16\pi G \int d^3y \;
{
\overline{T}^{\mu\nu}(\vec y,\tilde t) 
\over
|\vec x-\vec y|},
\end{equation}
where $\tilde t$ is the retarded time 
\begin{equation}
\tilde t = t - |\vec x-\vec y|. 
\end{equation}
These are the gravitational analog of the {\Lienard}--Wiechert
potentials of ordinary electromagnetism, and the integral has support
on the unperturbed backward light cone from the point $\vec x$. 

In writing down this formal solution we have tacitly assumed that
there is no incoming gravitational radiation. We have also assumed
that the \emph{global geometry} of spacetime is approximately
Minkowski, a somewhat more stringent condition than merely assuming
that the metric is \emph{locally} approximately Minkowski. Finally note
that the fact that we have been able to completely gauge-fix Einstein
gravity in a canonical manner is essential to argument. That we can
locally gauge-fix to the Hilbert--Lorentz gauge is automatic. By the
assumption of asymptotic flatness implicit in linearized Einstein
gravity, we can apply this gauge at spatial infinity where the only
remaining ambiguity, after we have excluded gravitational radiation,
is that of the Poincare group. (That is: Solutions of the
Hilbert--Lorentz gauge condition, which can be rewritten as $\nabla^2
x^\mu = 0$, are under these conditions unique up to Poincare
transformations.) We now extend the gauge condition inward to cover
the entire spacetime, the only obstructions to doing so globally
coming from black holes or wormholes, which are excluded by definition
of linearized gravity. Thus adopting the Hilbert--Lorentz gauge in
linearized gravity allows us to assign a \emph{canonical} flat
Minkowski metric to the entire spacetime, and it is the existence of
this canonical flat metric that permits us to make the comparisons
(between two different metrics on the same spacetime) that are at the
heart of the argument that follows.

Now consider a vector $k^\mu$ which we take to be a null vector of the
\emph{unperturbed\,} Minkowski spacetime
\begin{equation}
\eta_{\mu\nu} \; k^\mu k^\nu = 0.
\end{equation}
In terms of the full perturbed geometry this vector has a norm
\begin{eqnarray}
||k||^2 &\equiv&
g_{\mu\nu} \; k^\mu k^\nu 
\\
&=&
h_{\mu\nu} \; k^\mu k^\nu
\\
&=& 16\pi G \int d^3y \;
{ 
T_{\mu\nu}(\vec y,\tilde t)  \; k^\mu k^\nu
\over
|\vec x-\vec y|}.
\end{eqnarray}
Now assume the NEC
\begin{equation}
T_{\mu\nu} \; k^\mu k^\nu \geq 0, 
\end{equation}
and note that the kernel $|\vec x-\vec y|^{-1}$ is positive definite.
Using the fact that the integral of a everywhere positive integrand is
also positive, we deduce $g_{\mu\nu} \; k^\mu k^\nu \geq 0$.  Barring
degenerate cases, such as a completely empty spacetime, the integrand
will be positive definite so that
\begin{equation}
g_{\mu\nu} \; k^\mu k^\nu > 0.
\end{equation}
That is, a vector that is null in the Minkowski metric will be
spacelike in the full perturbed metric. Thus the null cone of the
perturbed metric must everywhere lie inside the null cone of the
unperturbed Minkowski metric.

Because the light cones contract, the \emph{coordinate} speed of light
must everywhere decrease. (Not the \emph{physical} speed of light as
measured by local observers, as always in Einstein gravity, that is of
course a constant.) This does however mean that the time required for
a light ray to get from one spatial point to another must always
increase compared to the time required in flat Minkowski space. This
is the well-known Shapiro time delay, and we see two important points:
(1) to even define the delay (delay with respect to what?) we need to
use the flat Minkowski metric as a background, (2) the fact that in
the solar system it is always a delay, never an advance, is due to the
fact that everyday bulk matter satisfies the NEC.

(We mention in passing that the strong energy condition [SEC] provides
a somewhat stronger result: If the SEC holds then the proper time
interval between any two timelike separated events in the presence of
the gravitational field is always larger than the proper time interval
between these two events as measured in the background Minkowski
spacetime.)

Now subtle quantum-based violations of the NEC are known to
occur~\cite{Visser:ANEC}, but they are always small and are in fact
tightly constrained by the Ford--Roman quantum
inequalities~\cite{Ford-Roman,Pfenning-Ford}. There are also {\em
classical} NEC violations that arise from non-minimally coupled scalar
fields~\cite{Flanagan-Wald}, but these NEC violations require
Planck-scale expectation values for the scalar field.  (These
classical NEC violations can however lead to very exotic phenomena such
as traversable wormholes~\cite{Classical}.) Be that as it may, NEC
violations are never appreciable in a solar system or galactic
setting. (SEC violations are on the other hand relatively common. For
example: cosmological inflation, classical massive scalar fields,
\etc.)

{From} the point of view of warp drive physics, this analysis is
complementary to that of~\cite{Olum}, (and also to the comments by
Coule~\cite{Coule}, regarding energy condition violations and
``opening out'' the light cones).  Though the present analysis is
perturbative around Minkowski space, it has the advantage of
establishing a direct and immediate \emph{physical} connection between
FTL travel and NEC violations.

\section{STRONG FIELD GRAVITY}

Generalizing these ideas beyond the weak field perturbative regime is
rather tricky: To even define effective FTL one will need to compare
two metrics. (Just to be able to ask the question ``FTL with respect
to what?'').

For instance, if we work perturbatively around a general metric,
instead of perturbatively around the Minkowski metric, then adopting
the Hilbert--Lorentz gauge again lets us make unambiguous statements
comparing the light cones of two metrics that differ
infinitesimally. However, there are other complications which are
already immense: (1) the Laplacian in the linearized gravitational
equations must be replaced by the Lichnerowicz operator; (2) the Green
function for the Lichnerowicz operator need no longer be concentrated
\emph{on} the past light cone [physically, there can be back-scattering
from the background gravitational field, and so the Green function can
have additional support from \emph{within} the backward light cone];
and (3) the Green function need no longer be positive definite.

Indeed, even for perturbations around a Friedman--Robertson--Walker
(FRW) cosmology, the analysis is not easy~\cite{FRW-Ford}. Because
linearized gravity is \emph{not} conformally coupled to the background
the full history of the spacetime back to the Big Bang must be
specified to derive the Green function. Furthermore, from the
astrophysical literature concerning gravitational lensing it is known
that {\emph{voids}} (as opposed to over-densities) can sometimes lead
to a Shapiro time {\emph{advance}}
\cite{Advance1,Advance2,Advance3}. This is not in conflict with the
present analysis and is not evidence for astrophysical NEC
violations. Rather, because those calculations compare a inhomogeneous
universe with a void to a homogeneous FRW universe, the existence of a
time advance is related to a suppression of the density below that of
the homogeneous FRW cosmology.  The local speed of light is determined
by the local gravitational potential relative to the FRW
background. Voids cause an increase of the speed of photons relative
to the homogeneous background.

A more promising attack on the notion of strong-field FTL is via the
notion of the relaxed Einstein equations, in which the full nonlinear
metric is written as
\begin{equation}
\bar h^{\mu\nu} = \eta^{\mu\nu} - \sqrt{-g}\; g^{\mu\nu}.
\end{equation}
Again adopting the Hilbert--Lorentz gauge, the \emph{full} nonlinear
Einstein equations can be written in terms of the flat space Laplacian
in the exact form (see, \eg, \cite{Will-Wiseman})
\begin{equation}
\Delta \bar h^{\mu\nu} = -16\pi G \; \tau^{\mu\nu},
\end{equation}
where the effective stress-energy pseudo-tensor is
\begin{equation}
\tau^{\mu\nu} = \sqrt{-g} \; T^{\mu\nu} + T_{LL}^{\mu\nu} + S^{\mu\nu}.
\end{equation}
The effective stress-energy pseudo-tensor is a combination of the
ordinary stress-energy tensor, the Landau--Lifshitz pseudo-tensor
($T_{LL}$), and a certain combination of second derivatives of the
metric ($S$) which has the effect of ``correcting'' the
characteristics [the light cones] away from the flat space light cones
to those of the full curved spacetime geometry. If the effective
stress-energy pseudo-tensor satisfies the NEC with respect to the flat
metric $\eta^{\mu\nu}$, then as before we can argue that the light
cones will always contract.

This argument cannot be viewed as fully satisfying since (1) it
requires a technical assumption about the global existence of the
Einstein--Hilbert gauge (at the very least, on the domain of outer
communication), and (2) while there are good physical reasons to
assert that the stress-energy tensor of bulk everyday matter should
satisfy the NEC with respect to the true spacetime metric, it is much
less clear whether there is any particular reason to believe that the
effective stress-energy pseudo-tensor should satisfy a NEC with
respect to the background Minkowski metric induced by the assumed
global Einstein--Hilbert gauge.

An attempt at dealing with the strong-field situation by using the
notion of Scri (past and future null infinities) has become bogged
down in issues of considerable technical complexity.

\section{DISCUSSION}

This note argues that any form of FTL travel requires violations of
the NEC.  The perturbative analysis presented here is very useful in
that it demonstrates that it is already extremely difficult to even
get even started: Any perturbation of flat space that exhibits even
the slightest amount of FTL (defined as widening of the light cones)
must violate the NEC. Moving beyond perturbation theory is both subtle
and technically difficult, and we do not yet have a fully convincing
argument that applies in a non-perturbative setting.


\end{document}